\documentclass[a4paper,10pt]{article}
\usepackage[dvips]{graphicx}
\begin{document}
\title{Exact Shock Profile for the ASEP with Sublattice-Parallel Update}
\author{F H Jafarpour$^{1,2}$ \footnote{Corresponding author's e-mail:farhad@ipm.ir}, 
F E Ghafari$^{1}$ and S R Masharian$^{1}$ \\ \\
{\small $^1$Bu-Ali Sina University, Physics Department, Hamadan, Iran} \\
{\small $^2$Institute for Studies in Theoretical Physics and Mathematics (IPM),} \\
{\small P.O. Box 19395-5531, Tehran, Iran } }
\maketitle
\begin{abstract}
We analytically study the one-dimensional Asymmetric Simple
Exclusion Process (ASEP) with open boundaries under sublattice-parallel 
updating scheme. We investigate the stationary state properties 
of this model conditioned on finding a given particle number 
in the system. Recent numerical investigations have shown that 
the model possesses three different phases in this case. Using 
a matrix product method we calculate both exact canonical partition 
function and also density profiles of the particles in each phase. 
Application of the Yang-Lee theory reveals that the model undergoes two
second-order phase transitions at critical points. These results
confirm the correctness of our previous numerical studies.\\[5mm]
\end{abstract}

\section{Introduction}
The study of steady states of non-equilibrium systems has motivated 
a lot of works over the last decades \cite{lig,fr,sch-rev}. These 
models exhibit in general shock structures in their steady state. 
The study of microscopic and macroscopic structure of shocks in
one-dimensional reaction-diffusion models has also been one of
physicists' interests recently \cite{bs,dls,kjs,ps}. The most 
well known model which reveals shocks in the steady state is Asymmetric
Simple Exclusion Process (ASEP) with open boundaries \cite{sd,dehp}.
The ASEP is a model of diffusing identical classical particles
with hardcore interactions on a one-dimensional lattice. In this
model particles are injected from left boundary of a lattice of
length $L$ with rate $\alpha$ provided that the first site is
empty. They hop to right in the bulk of that lattice with finite
rate and leave it from right boundary with rate $\beta$ provided
that the last site is already occupied. A substantial amount of
exact results have been obtained for the ASEP which include the
phase diagram, stationary state probability distribution function,
correlation functions and correlation lengths. It is also known
that for special tuning of microscopic rates a travelling shock
with a step-like density profile evolves in the system. The phase
diagram of the ASEP with random sequential updating scheme
(continuous-time) has three different phases: a low-density phase,
a high-density phase and a maximal-current phase. The shocks
appear on the coexistence line between the low-density and
high-density phases, however; for the ASEP with open boundaries
the shock can be anywhere on the lattice with equal probabilities
and the resulting profile in the steady state is linear. In order
to study the dynamics of the shocks in the ASEP, the model been
considered on a lattice with periodic boundary condition in the
presence of a second-class particle (often called an impurity)
\cite{fr,dls,djls}. In this case the density profile of the 
particles, as seen from the impurity, is a step-like function i.e. a
discontinuity between a low-density and a high-density region with
an interface which is sharp on lattice scale.\\
Despite the remarkable work on the microscopic structure of shocks 
and their dynamics in continuous-time update \cite{bs,dls,kjs}, 
not much has been known about that in discrete-time updates. 
For the ASEP in sublattice-parallel update the temporal
evolution of a single shock has been studied \cite{ps}. By making
use of a quantum algebra symmetry of the generator of the process
it has been shown that on the microscopic level the dynamics of
the shock can be described by a single particle dynamics in the
thermodynamic limit. In the present paper we study the microscopic
structure of the shocks in the ASEP in sublattice-parallel
dynamics; however, instead of introducing a second-class particle,
we consider the case where the total number of particles is kept fixed.\\
The ASEP under sublattice-parallel update with open boundaries was 
first introduced and studied in \cite{sch}. In this model particles 
are injected from the left boundary with probability $\alpha$ and 
extracted from the right boundary with probability $\beta$. In the 
bulk of the lattice they hop to the right with unit probability. The 
exact phase diagram and stationary density profile of the particles 
in this case (without conditioning on the particle number) have been obtained 
in the same reference. Later this model was studied using a matrix 
product method and the same results were obtained \cite{hin}. However,
the case of fixed particle number has not been studied yet.\\
Our recent numerical investigations show that in the case of fixed particle number
the phase diagram of the model highly depends on the total density of the
particles on the lattice $\rho$, $\alpha$ and $\beta$ \cite{f}.
For $\rho < \frac{1}{2}$ the phase diagram contains a low-density
and a shock phase while for $\rho > \frac{1}{2}$ it has a
high-density phase and also a shock phase. In both cases the phase
transition points $(\alpha_c,\beta_c)$ are determined by the total
density of particles $\rho$.  The density profile of the particles
in the steady state have also been studied for both even and odd
sites. In low-density (high-density) phase $\rho < \frac{1}{2},
\beta > \beta_c$ ($\rho > \frac{1}{2}, \alpha > \alpha_c$) the
density profile of the particles for both even and odd sites is a
constant on the lattice except near the left (right) boundary
where it becomes an exponentially increasing (decreasing) function
of site number. In the shock phase $\rho < \frac{1}{2}, \beta <
\beta_c$ or $\rho > \frac{1}{2}, \alpha < \alpha_c$ the density
profile of the particles for both even and odd sites is an error function.\\
In this paper we continue our studies analytically. Using a matrix
product formalism, in which the steady state weights are written
in terms of the expectation value of product of non-commuting
operators, we calculate the canonical partition function of the
model exactly. The thermodynamic behavior of this function will
also be calculated in each of three phases mentioned above. The
exact expressions for the density profile of the particles in each
phase have also been calculated. Recently it has been shown that
the classical Yang-Lee theory of equilibrium phase transitions
\cite{yl} can be applied to the non-equilibrium models to study
their out of equilibrium phase transitions (for a review see
\cite{eb}). The simplicity of our model allows us to calculate the
line of the canonical partition function zeros in the
thermodynamic limit from which we obtain the critical points. In
analogy with equilibrium statistical mechanics one can introduce
the pressure (which is obviously not the physical pressure of the
particles) and study its behavior as a functions of $\rho^{-1}$
for fixed $\alpha$ and $\beta$. It turns out that resulting curve
looks quite similar to the isotherm of a Bose gas when Bose-Einstein 
condensation takes place.\\
Our paper is organized as follows: In section 2 we will define the
model and calculate its steady state probability distribution
function using a matrix product formalism. The canonical partition
function of the model will also be calculated. In section 3 the
density profile of the particles on the lattice will be obtained.
In section 4 we will apply the Yang-Lee theory to study the phase
transitions in the model and also study its similarities with 
a Bose gas. Finally in section 5 we will discuss the results.
\section{Canonical partition function}
\begin{figure}[htbp]
\includegraphics[height=5cm] {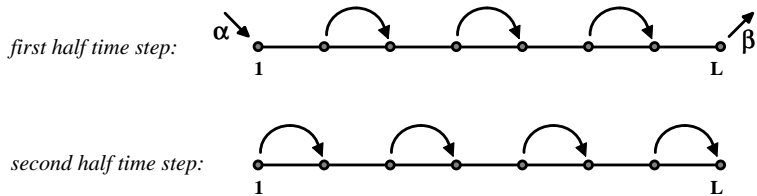}
\caption{\label{fig1} The ASEP under sublattice-parallel updating scheme.}
\end{figure}

\begin{figure}[htbp]
\begin{center}
\includegraphics[width=11cm,height=3.7cm] {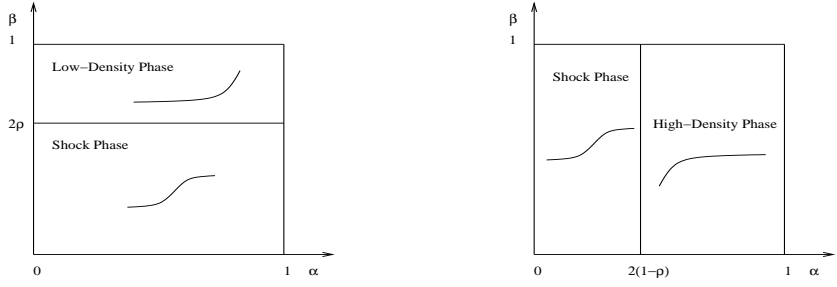}
\caption{\label{fig2} The phase diagrams of the model for $\rho <
\frac{1}{2}$ (left) and $\rho > \frac{1}{2}$ (right). The small
curves show the behaviors of the density profile of particles in
each phase obtained from numerical calculations.}
\end{center}
\end{figure}

The ASEP with sublattice-parallel update is defined as follows:
Classical particles hop deterministically on a one-dimensional
lattice of length $L$ which is assumed to be an even number. The
system is updated in two half time steps. In the first half time
step the leftmost and the rightmost sites and also the particles
at even sites are updated. They are injected to the leftmost site
with probability $\alpha$ provided that the first site is empty.
The particles at even sites will hop to their rightmost neighbors
provided that the target sites are empty. If the last site of the
lattice is occupied it will be extracted with probability $\beta$.
In the second half-time step only the particles at odd sites will
be updated according to the bulk dynamics mentioned above. These
two steps are shown in Fig.~\ref{fig1}.
From now on we assume that the total density of particles on the lattice 
is kept fixed $\rho=\frac{M}{L}$ in which $M$ is the total number of particles. 
In order to study the steady state properties of this model analytically we 
will adopt a so called Matrix Product Formalism (MPF). In what follows we will 
briefly review the MPF for the ASEP with sublattice-parallel dynamics
first introduced in \cite{hin}. Let us define the occupation
number at site $i$ as $\tau_i \in \{0,1\}$ where $\tau_i=1$ if it
is occupied and $\tau_i=0$ if it is empty. Now according to the
MPF, the stationary state probability for the system to be in a
given configuration $\{ \tau \}=\{ \tau_1, \cdots, \tau_L\}$ with
exactly $M$ particles is given by
\begin{equation}
\label{MA}
P(\{ \tau \}) \propto \delta (M-\sum_{i=1}^{L}\tau_i) \langle W \vert
\prod_{i=1}^{\frac{L}{2}} [\hat{\mathcal{O}}_{2i-1}\mathcal{O}_{2i}]
\vert V \rangle
\end{equation}
where we have defined
\begin{equation}
\label{operators}
\hat{\mathcal{O}}_i:=\tau_{i} \hat{D} + (1-\tau_{i})\hat{E} \; , \;
\mathcal{O}_i:=\tau_{i} D + (1-\tau_{i})E.
\end{equation}
The Dirac $\delta$ in (\ref{MA}) guarantees the conservation of
the total number of particles. For the ASEP with
sublattice-parallel dynamics the operators ($\hat{D},\hat{E}$) and
($D,E$) are finite dimensional square matrices and with the vectors 
$\vert V \rangle$ and $\langle W \vert$ satisfy the following quadratic 
algebra
\begin{equation}
\begin{array}{l}
\label{BulkAlgebra}
[E,\hat{E}] = [D,\hat{D}] = 0 \; ,
\; E \hat{D} = [\hat{E},D] \; , \; \hat{D} E = 0 \\
\langle W \vert \hat{E} (1-\alpha) = \langle W \vert E \; , \;
\langle W \vert (\alpha \hat{E} + \hat{D}) = \langle W \vert D \\
(1-\beta) D \vert V \rangle = \hat{D} \vert V \rangle \; , \;
(E+\beta D) \vert V \rangle = \hat{E}\vert V \rangle.
\end{array}
\end{equation}
The operators ($\hat{D},\hat{E}$) and ($D,E$) stand for the
presence of particles and holes at odd and even sites
respectively. It has also been shown that (\ref{BulkAlgebra}) has
a two-dimensional representation for $\alpha \neq \beta$ given by
\cite{hin}
\begin{equation}
\label{Representation}
\begin{array}{l}
\label{BulkRep} \hat{D} = \left( \begin{array}{cccc}
\alpha (1-\beta) & & & 0 \\
-\alpha \beta & & & 0
\end{array} \right) \; , \;
\hat{E} = \left( \begin{array}{cccc}
\alpha \beta & & & 0  \\
\alpha\beta & & & \beta
\end{array} \right) \; ,\;
\vert V \rangle = \left( \begin{array}{c}
1-\beta \\ -\beta \end{array} \right) \\ \\
D = \left( \begin{array}{cccc}
\alpha  & & & 0 \\
-\alpha \beta & & & \alpha \beta
\end{array} \right) \; , \;
E = \left( \begin{array}{cccc}
0 & & & 0  \\
\alpha\beta & & & \beta (1-\alpha)
\end{array} \right) \; , \;
\langle W \vert = (\alpha, 1-\alpha).
\end{array}
\end{equation}
The normalization factor in (\ref{MA}), which will be called the
canonical partition function of the model, should be obtained
using the fact that the steady state probability distribution
function $P(\{ \tau \})$ is normalized i.e. $\sum_{\{ \tau \}}
P(\{ \tau \})=1$. It can now be written as
\begin{equation}
Z_{L,M}=\sum_{\{\tau\}}\delta(M-\sum_{i=1}^{L}\tau_i)\langle W \vert
\prod_{i=1}^{\frac{L}{2}} [\hat{\mathcal{O}}_{2i-1}\mathcal{O}_{2i}]
\vert V \rangle.
\end{equation}
The canonical partition function of the model $Z_{L,M}$ is a key
function in calculating the physical quantities such as the 
density profile of the particles on the lattice in the steady
(which will be discussed in next section), therefore; we
will first concentrate on calculating this quantity. As we will
see it is more easier to calculate so called grand canonical
partition function of the model which is defined as
\begin{eqnarray}
\label{GCPF} Z_{L}(\xi) & = & \sum_{\{\tau \}} \langle W
\vert\prod_{i=1}^{\frac{L}{2}} (\tau_{2i-1} \xi {\hat
D}+(1-\tau_{2i-1}){\hat E})(\tau_{2i}\xi D+(1-\tau_{2i})E)\vert V
\rangle \nonumber \\ & = & \langle W \vert ({\hat C}
C)^{\frac{L}{2}} \vert V \rangle=\sum_{M=0}^{L}\xi^{M} Z_{L,M}
\end{eqnarray}
where $\xi$ is the fugacity associated with the particles and
also we have defined $\hat C:=\xi{\hat D}+ {\hat E}$ and $C:=\xi
D+E$. The total density of particles on the lattice $\rho$ will
then be fixed by the fugacity of them $\xi$ through the following
relation
\begin{equation}
\label{DFR}
\rho=\lim_{L\rightarrow\infty}\frac{\xi}{L}\frac{\partial}{\partial
\xi} \ln Z_L(\xi).
\end{equation}
Now one can easily calculate the canonical partition function of
the system by inverting the series in (\ref{GCPF}) using
\begin{equation}
\label{IR} Z_{L,M}=\frac{1}{2\pi i}\int_{C}d\xi\frac{Z_L(\xi)}{\xi^{M+1}}
\end{equation}
where $C$ is a contour which encircles the origin anti-clockwise. \\
Let us start with calculating the grand canonical partition function
of the model using (\ref{Representation}) and (\ref{GCPF}).
This can easily be done and one finds
\begin{equation}
Z_{L}(\xi)=\langle W \vert ({\hat C} C)^{\frac{L}{2}} \vert V
\rangle=Z^{(1)}_L (\xi)+Z^{(2)}_L (\xi)
\end{equation}
in which we have defined
\begin{eqnarray}
Z^{(1)}_L (\xi) & = & -\frac{\xi \alpha (\alpha-\beta)(1-\beta)}
{\beta(1-\alpha)-\xi\alpha(1-\beta)}u_1^{\frac{L}{2}}\\
Z^{(2)}_L (\xi) & = & \frac{\beta(\alpha-\beta)(1-\alpha)}
{\beta(1-\alpha)-\xi\alpha(1-\beta)}u_2^{\frac{L}{2}},
\end{eqnarray}
where $u_1=\xi \alpha^2(\beta+\xi(1-\beta))$ and $u_2=\beta^2(\xi
\alpha+(1-\alpha))$. In the thermodynamic limit $L \rightarrow
\infty$ one can easily distinguish three different cases
associated with three different phases:
\begin{itemize}
\item[$\bullet$]Case $u_1 >u_2$\\
In this case we have $Z_L(\xi) \simeq Z^{(1)}_L(\xi)$. Using
(\ref{DFR}) one finds $\xi=\frac{\beta(2
\rho-1)}{2(1-\beta)(1-\rho)}$ in this phase. For the fugacity to
be positive we should have $\rho > \frac{1}{2}$. Now the condition
$u_1 >u_2$ translates to $\alpha > 2(1-\rho)$. These two
conditions determine the boundaries of this phase which will be
called {\em High-Density} phase hereafter. In order to calculate
the canonical partition function of the model in high-density
phase we apply the standard steepest descent method to (\ref{IR}).
After some calculations we find the following expression for the
canonical partition function of the model in this phase:
\begin{equation}
\label{CPF1}
Z_{L,M} \simeq \frac{(\alpha-\beta)\alpha^{L+1}\beta^{L-M}(2\rho-1)^{\frac{L}{2}
-M+\frac{1}{2}}}{(\alpha-2(1-\rho))(1-\beta)^{\frac{L}{2}-M}(2(1-\rho))^{L-M+\frac{1}{2}}}
\end{equation}
As we will see the density profile of the particles in this phase
has an exponential behavior near the left boundary and is a
constant elsewhere.
\item[$\bullet$]Case $u_1 <u_2$\\
In this case we have $Z_L(\xi) \simeq Z^{(2)}_L(\xi)$. By using
(\ref{DFR}) one finds
$\xi=\frac{2\rho(1-\alpha)}{\alpha(1-2\rho)}$ in this phase. As
before the fugacity should be positive, therefore; we should have
$\rho < \frac{1}{2}$. The condition $u_1 < u_2$ also translates to
$\beta > 2 \rho$. These two conditions i.e. $\rho < \frac{1}{2}$
and $\beta > 2 \rho$ determine the boundaries of the current
phase. Using the standard steepest descent method we find the
following expression for the canonical partition function of the
model in this phase which will be called {\em Low-Density} phase
hereafter:
\begin{equation}
\label{CPF2}
Z_{L,M} \simeq \frac{(\alpha-\beta)\alpha^M\beta^{L+1}(1-\alpha)^{\frac{L}{2}-M}}
{(\beta-2\rho)(2\rho)^{M+\frac{1}{2}}(1-2\rho)^{\frac{L}{2}-M-\frac{1}{2}}}
\end{equation}
As we will see the density profile of the particles in this phase
has an exponential behavior near the right boundary and is a
constant elsewhere.
\item[$\bullet$]Case $u_1=u_2$\\
The remaining phase corresponds to the regions $\rho <
\frac{1}{2},\beta < 2 \rho$ and $\rho > \frac{1}{2},\alpha < 2
(1-\rho)$. The condition $u_1=u_2$ gives
$\xi_0=\frac{\beta(1-\alpha)}{\alpha(1-\beta)}$. It can easily be
seen that both $Z^{(1)}_L$ and $Z^{(2)}_L$ have a pole at this
point. One can also verify that the density-fugacity relation
(\ref{DFR}) fails in this phase i.e. the fugacity cannot fix the
total density of particles on the lattice. It is known that the
physical meaning for this is the fact that shocks appear in this
phase. This is the reason that we will call it the {\em Shock}
phase hereafter. In order to calculate the canonical partition
function of the model in this phase we will again apply the
steepest descent method. Let us take the contour of the integral a
circle with radius $R$ around the origin. The saddle point
associated with $Z^{(1)}_L$ will be
$\xi_1=\frac{\beta(2\rho-1)}{2(1-\beta)(1-\rho)}$ and for
$Z^{(2)}_L$ it will be
$\xi_2=\frac{2\rho(1-\alpha)}{\alpha(1-2\rho)}$. In the shock
phase we have $\xi_1 < \xi_0 < \xi_2$. First we take the contour
of the integral $R$ as $R=\xi_2$ for both $Z^{(1)}_L$ and
$Z^{(2)}_L$. Then one can modify the contour of the integral from
$R=\xi_2$ to $R=\xi_1$ for $Z^{(1)}_L$. Since $Z^{(1)}_L$ has a
pole at $\xi_0$ there is a contribution from the pole which can be
calculated using the cauchy residue theorem. The result is that
the residue $Z^{(1)}_L$ at $\xi_2$ is the largest contribution to
the canonical partition function of the model and therefore we
have
\begin{equation}
\label{CPF3}
Z_{L,M} \simeq (\alpha-\beta) \alpha^M \beta^{L-M}
(\frac{1-\alpha}{1-\beta})^{\frac{L}{2}-M}.
\end{equation}
\end{itemize}
In Fig.~\ref{fig2} we have plotted the phase diagram of the model
obtained from above discussion. One should note that the dynamical rules
in the present model are exactly the ones in \cite{sch}; however,
because of conditioning on the density of particles we find 
a different phase diagram.
\section{Density profile of the particles}
In this section we will define and calculate the exact expressions
for the density profiles of the particles on the lattice in the
steady state. As for the partition function it is much easier to
calculate the unnormalized average particle number in the grand
canonical ensemble and then translate the results into those in
the canonical ensemble using the inversion formula
\begin{equation}
\label{IRDP} \langle \rho_i \rangle_{L,M}^{(u)} = \frac{1}{2 \pi i}
\int_{C}d\xi \frac{\langle \rho_i\rangle_{L}^{(u)}(\xi)}{\xi^{M+1}}.
\end{equation}
The normalized average particle number at site $i$ should then be
obtained from
\begin{equation}
\label{NCDP}
\langle \rho_i \rangle=\langle \rho_i \rangle_{L,M}^{(u)}/Z_{L,M}.
\end{equation}
The average particle number on even sites in the grand canonical
ensemble is defined as
\begin{eqnarray}
\label{DPE1} \langle \rho_{2i}\rangle_L^{(u)}(\xi) = \langle W
\vert (\hat{C}C)^{i-1}\hat{C}\xi D (\hat{C}C)^{\frac{L}{2}-i}
\vert V \rangle \;\;,\;\; 1 \leq i \leq \frac{L}{2}
\end{eqnarray}
and for odd sites in the same ensemble
\begin{eqnarray}
\label{DPO1} \langle \rho_{2i-1}\rangle_L^{(u)}(\xi) = \langle W
\vert (\hat{C}C)^{i-1} \xi \hat{D} C (\hat{C}C)^{\frac{L}{2}-i}
\vert V \rangle  \;\;,\;\; 1 \leq i \leq \frac{L}{2}
\end{eqnarray}
in which $\hat{C}$ and $C$ have the same previous definitions.
Using the representation of the quadratic algebra
(\ref{BulkAlgebra}) given by (\ref{Representation}) and after some
algebra one can calculate (\ref{DPE1}) and (\ref{DPO1}) to find
\begin{eqnarray}
\label{DPE2}
\langle \rho_{2i}\rangle_L^{(u)}(\xi) & = & -\frac{\xi \alpha (\alpha-\beta)(1-\beta)}
{\beta(1-\alpha)-\xi\alpha(1-\beta)}u_1^{\frac{L}{2}}+\\
& & \frac{\xi \alpha \beta (1-\alpha)(\alpha-\beta)}{(1+(\xi-1)\alpha)(\beta(1-\alpha)
-\xi\alpha(1-\beta))} u_2^{\frac{L}{2}}+\nonumber \\
& & \frac{\xi \alpha (\alpha-\beta)(1-\alpha)(1-\beta)}{(1+(\xi-1)\alpha)(\beta(1-\alpha)
-\xi\alpha(1-\beta))}u_2^i u_1^{\frac{L}{2}-i} \nonumber
\end{eqnarray}
and
\begin{eqnarray}
\label{DPO2} \langle \rho_{2i-1}\rangle_L^{(u)}(\xi) & = &
-\frac{\xi^2 \alpha (\alpha-\beta)(1-\beta)^2}{
(\beta+\xi(1-\beta))(\beta(1-\alpha)-\xi\alpha(1-\beta))} u_1^{\frac{L}{2}} \\
& & +
\frac{\xi^2\alpha^2(1-\alpha)(\alpha-\beta)(1-\beta)}{\beta(1+(\xi-1)
\alpha)(\beta(1-\alpha)-\xi\alpha(1-\beta))}u_2^i
u_1^{\frac{L}{2}-i}. \nonumber
\end{eqnarray}
In what follows we will study the thermodynamic behaviors of
(\ref{DPE2}) and (\ref{DPO2}) in each of the above mentioned
phases i.e. the high-density, low-density and the shock phase.
\begin{itemize}
\item[$\bullet$]The high-density phase  ($\rho > \frac{1}{2}, \alpha > 2(1-\rho)$) \\
In this phase we have $u_1 > u_2$, therefore; in the thermodynamic
limit the dominant terms in (\ref{DPE2}) are the first and the
last term, however; we should keep both the first and the second
terms in (\ref{DPO2}) in the thermodynamic limit in this phase.
Using (\ref{CPF1}), (\ref{IRDP}), (\ref{NCDP}) and by applying the
steepest descent method after some algebra one finds
\begin{eqnarray}
& & \langle \rho_{2i}\rangle   =  1-\frac{2(1-\alpha)(1-\beta)(1-\rho)}
{2(1-\beta)(1-\rho)+\alpha(\beta-2(1-\rho))}e^{-\frac{i}{\zeta}} \\
& & \langle \rho_{2i-1}\rangle  = (2\rho-1)-\frac{\alpha(1-\alpha)(2\rho-1)}
{2(1-\beta)(1-\rho)+\alpha(\beta-2(1-\rho))}e^{-\frac{i}{\zeta}}
\end{eqnarray}
in which $i=1, \cdots, \frac{L}{2}$ and that we have defined the
correlation length $\zeta^{-1}=\vert
\ln\frac{2(1-\rho)(2(1-\beta)(1-\rho)+\alpha(\beta-2(1-\rho)))}{\alpha^2
(2\rho-1)} \vert$. As can be seen both these density profiles are
exponentially increasing functions of $i$ near the left boundary,
however; as we go farther from there they remain constant in the
bulk of the lattice and also near the right boundary.
\item[$\bullet$]The low-density phase ($\rho < \frac{1}{2}, \beta > 2\rho$) \\
In this phase we have $u_1 < u_2$, therefore; in the thermodynamic
limit we should keep the second and the third terms in
(\ref{DPE2}) and only the second term in (\ref{DPO2}). Using
(\ref{CPF1}), (\ref{IRDP}), (\ref{NCDP}) and by applying the
steepest descent method after some algebra one finds
\begin{eqnarray}
& & \langle \rho_{2i}\rangle   =  2\rho+\frac{2\rho(1-\beta)}{\beta}
e^{-\frac{\frac{L}{2}-i}{\zeta}} \\
& & \langle \rho_{2i-1}\rangle  =  \frac{4(1-\alpha)(1-\beta)\rho^2}
{\beta^2(1-2\rho)}e^{-\frac{\frac{L}{2}-i}{\zeta}}
\end{eqnarray}
in which $i=1, \cdots, \frac{L}{2}$ and we have defined
the correlation length $\zeta^{-1}=\vert
\ln\frac{2\alpha\beta\rho-4(\alpha+\beta-1)\rho^2}{\beta^2(1-2\rho)}
\vert$. It can be seen that the density profiles of the particles
at both even and odd sites are constant near the left boundary and
also in the bulk of the lattice while they are exponentially
increasing functions of $i$ near the right boundary.
\item[$\bullet$]The shock phase ($\rho > \frac{1}{2}, \alpha < 2(1-\rho)$ and
$\rho < \frac{1}{2}, \beta < 2\rho$) \\
As we mentioned above the density-fugacity relations fails in this
phase and it can be a sign for the existence of shocks in the
density profile of the particles on the lattice. The density of
particles at even sites on the left hand side of the shock
(low-density region) is $\rho_{low-even}=\beta$ and on the right
hand side of the shock (high-density region) is
$\rho_{high-even}=1$. These values for the density of particles at
odd sites are $\rho_{low-odd}=0$ and $\rho_{high-odd}=1-\alpha$
respectively. In what follows we will investigate the shock width
in more details. In order to calculate the density profile of the
particles in this phase we adopt the following procedure: for
large system size the density profile of particles on the lattice
can be described by a continuous function $\rho(x)$ in which
$x=\frac{i}{L}$ and $0 \leq x \leq 1$. From (\ref{DPE2}) and
(\ref{DPO2}) one obtains
\begin{eqnarray}
& & \langle \rho_{2i+2}\rangle_L^{(u)}(\xi) - \langle \rho_{2i} \rangle_L^{(u)}(\xi)
\propto u_2^{i} u_1^{\frac{L}{2}-i} \\
& & \langle \rho_{2i+1}\rangle_L^{(u)}(\xi) - \langle \rho_{2i-1} \rangle_L^{(u)}(\xi)
\propto u_2^{i} u_1^{\frac{L}{2}-i}.
\end{eqnarray}
Now it can easily be verified that for the density of particles at even and odd sites we have
\begin{equation}
\frac{d}{dx} \rho_{even/odd}(x,\xi)  =  \rho_{even/odd}(\xi)  e^{L \cdot F(x,\xi)}
\end{equation}
in which we have defined $F(x,\xi)=(\frac{1}{2}-x)\ln u_1 + x \ln
u_2$. Using (\ref{IRDP}) and by applying the steepest descent
method we find
\begin{equation}
\frac{d}{dx} \rho_{even/odd}(x)  =  \rho_{even/odd}  e^{L \cdot G(x)}
\end{equation}
where $G(x)=(F(x,\xi)-\rho\ln\xi)\vert_{\xi=\xi_0}$ and $\xi_0$ is
the saddle point of the integral. It can be shown that $G(x)$ has
its maximum value at
$x_0=\frac{2(1-\rho)-\alpha}{2(2-\alpha-\beta)}$, therefore; one
can expand it around $x_0$ up to the second order to find
\begin{equation}
\frac{d}{dx} \rho_{even/odd}(x) =  \rho_{even/odd}  e^{\frac{L}{2} \cdot
G''(x_0) \cdot (x-x_0)^2}
\end{equation}
in which
\begin{equation}
G''(x_0)=\frac{-2(2-\alpha-\beta)^3}{\beta^2(\alpha-2(1-\rho))+\beta((\alpha-1)^2+
(1-2\rho))+2\alpha\rho(1-\alpha)}.
\end{equation}
By integrating this expression and applying the above mentioned
boundary conditions we find the following expressions for the
density profile of the particles at even and odd sites in the
shock phase
\begin{eqnarray}
& &  \rho_{even}(x)=\frac{1+\beta}{2}+\frac{1-\beta}{2}
erf(\sqrt{\frac{L}{2} \vert G''(x_0)\vert }(x-x_0)) \\
& &  \rho_{odd}(x)=\frac{1-\alpha}{2}+\frac{1-\alpha}{2}
erf(\sqrt{\frac{L}{2} \vert G''(x_0) \vert }(x-x_0))
\end{eqnarray}
where $erf(\cdots)$ is the error function. From this it follows
that the position of the microscopic shock position fluctuates 
around its mean value $x_0$ in a region of size $L^{1/2}$.
\end{itemize}
\section{Yang-Lee zeros and similarities with a Bose gas}
In this section we will first study the applicability of the classical
Yang-Lee theory to predict the phase transitions and their orders
in our model. According to this theory the zeros of the canonical
partition function as a function of one its intensive variables
lie on a curve which might interest the real positive axis of that
variable in the thermodynamic limit in a couple of points
\cite{yl}. In this case the system undergoes a phase transition at
these points. The order of transition can also be obtained by
investigating the angle at which the line of the partition
function zeros intersects the real positive axis. If the angle is
$\frac{\pi}{2n}$ then $n$ will be the order of transition. By
defining the free energy of the system as
$g=\lim_{L,M\rightarrow\infty}\frac{1}{L} \ln Z_{L,M}$ the line of
the canonical partition function zeros can be obtained using
\cite{gr}
\begin{equation}
\label{LOZ}
Re(g_1-g_2)=0
\end{equation}
in which $g_1$ and $g_2$ are free energies of the system on the
left and right hand side of the transition point. One can make use
of (\ref{CPF1})-(\ref{CPF3}) ,(\ref{LOZ}) and also the definition
of the free energy to obtain the line of partition function zeros
for transition between low-density and shock phases and also
between high-density and shock phase. For the transition between
high-density and shock phases after some algebra one finds
\begin{eqnarray}
\label{LOZ1}
\frac{(u^2+v^2)^{\frac{\rho-1}{2}}}{((1-u)^2+v^2)^{\frac{\rho}{2}-\frac{1}{4}}}=
\frac{(2(1-\rho))^{\rho-1}}{(2\rho-1)^{\rho-\frac{1}{2}}}
\end{eqnarray}
in which we have defined $\alpha=u+iv$. It can readily be verified
that this curve intersects the real positive $\alpha$ axis at
$\alpha_c=2(1-\rho)$ at an angle $\frac{\pi}{4}$; therefore, the
phase transition is of second-order. On the other hand for the
transition between low-density and shock phases we find
\begin{eqnarray}
\label{LOZ2}
\frac{(x^2+y^2)^{\frac{\rho}{2}}}{((1-x)^2+y^2)^{\frac{\rho}{2}-\frac{1}{4}}}=
\frac{(2\rho)^{\rho}}{(1-2\rho)^{\rho-\frac{1}{2}}}
\end{eqnarray}
in which we have defined $\beta=x+iy$. 
\begin{figure}[htbp]
\begin{center}
\includegraphics[height=2.9cm] {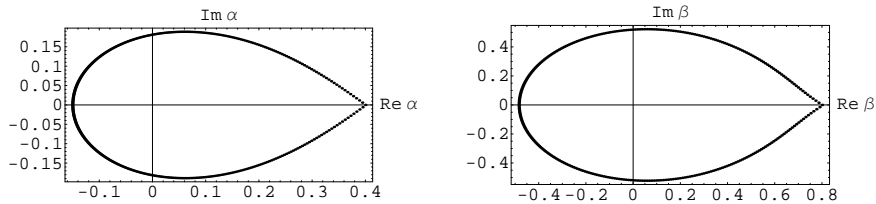}
\caption{\label{fig3} The line of the canonical partition function
zeros (\ref{LOZ1}) (left) and (\ref{LOZ2}) (right) for $\rho=0.8$
($\alpha_c=0.4$) and $\rho=0.4$ ($\beta_c=0.8$) respectively.}
\end{center}
\end{figure}
This curve also intersects the real positive $\beta$ axis at 
$\beta_c=2 \rho$ at an angle $\frac{\pi}{4}$; therefore, as 
before a second-order phase transition takes place. 
In Fig.~\ref{fig3} we have plotted the line of the zeros 
obtained from (\ref{LOZ1}) and (\ref{LOZ2}).\\
Let us now study the similarity between the phase transition in 
our model and that in a Bose gas. In order to see this similarity 
let us define pressure in terms of the grand canonical partition 
function of the model as $P=\frac{1}{L} \ln Z_L(\xi)$. 
Using the asymptotic behavior of the
grand canonical partition function of the model in the
thermodynamic limit and (\ref{DFR}) one can calculate the pressure
as a function of total density of particles $\rho$. This can
easily be done and we find
\begin{equation}
\label{P}
P=\cases{
\frac{1}{2}\ln (\frac{\alpha^2 \beta^2 (2\rho-1)}{4(1-\beta)(1-\rho)^2}),
&$\rho > \frac{1}{2},\alpha > 2(1-\rho)$ \cr
\frac{1}{2}\ln (\frac{\beta^2(1-\alpha)}{1-2\rho}),&$\rho < \frac{1}{2},
\beta > 2\rho$ \cr
\frac{1}{2}\ln (\frac{\beta^2(1-\alpha)}{1-\beta}),&$\rho > \frac{1}{2},
\alpha < 2(1-\rho) \; \; \mbox{and} \; \; \rho < \frac{1}{2},\beta < 2\rho.$ \cr }
\end{equation}
As can be seen the pressure as a function of density remain a
constant in the shock phase. It is also a decreasing function of
$\frac{1}{\rho}$ in both low-density and high-density phases. This
spectacular behavior reminds us the isotherm of a Bose gas when a
Bose-Einstein condensation takes place.
\section{Conclusion}
In this paper we studied the ASEP with sublattice-parallel update
and open boundaries where the mean particle number is kept fixed. 
Exact calculations using the MPF show that the system undergoes two
second-order phase transitions. The shock phase exists both for
$\rho > \frac{1}{2}$ and $\rho < \frac{1}{2}$ depending on the
values of $\alpha$ and $\beta$. For $\rho=\frac{1}{2}$ one has
only shock-phase regardless of injection and extraction
probabilities. In the shock phase the density profile of the
particles is an error function and its center is located at
$i_0=\frac{2(1-\rho)-\alpha}{2(2-\alpha-\beta)}L$. \\
In order to study the shock dynamics one can consider 
the ASEP with sublattice-parallel update on a ring in the 
presence of a second class particle. Our procedure can also 
be applied to calcuate the shock profiles in the partially 
ASEP (PASEP) with open boundaries under sublattice-parallel update \cite{hp}. 
\section{References}

\end{document}